\documentclass[aps,prb,superscriptaddress,twocolumn,showpacs]{revtex4}
\usepackage{graphics}
\usepackage[caption=false]{subfig}
\usepackage{epsfig}
\usepackage{color}
\usepackage{amsmath}
\usepackage{amssymb}
\usepackage{amsfonts}
\usepackage{wrapfig}
\usepackage{pstricks}
\usepackage{pst-node}
\usepackage{bm}
\usepackage{dcolumn}% Align table columns on decimal point
\usepackage{multirow}

\begin{document}
\title{Polarization analysis of excitons in monolayer and bilayer transition-metal dichalcogenides}

\author{Hanan~Dery}
\altaffiliation{hanan.dery@rochester.edu}
\affiliation{Department of Electrical and Computer Engineering, University of Rochester, Rochester, New York 14627, USA}
\affiliation{Department of Physics and Astronomy, University of Rochester, Rochester, New York 14627, USA}

\author{Yang~Song}
\affiliation{Department of Electrical and Computer Engineering, University of Rochester, Rochester, New York 14627, USA}

\begin{abstract}
The polarization analysis of optical transitions in monolayer and bilayer transition-metal  dichalcogenides provides invaluable information on the spin and valley (pseudospin) degrees of freedom. To explain optical properties of a given monolayer transition-metal  dichalcogenide, one should consider (i) the order of its spin-split conduction bands,  (ii) whether intervalley scattering is prone to phonon bottleneck, (iii) and whether valley mixing by electron-hole exchange can take place.  Using these principles, we present a consistent physical picture that elucidates a variety of features in the optical spectra of these materials. We explain the differences between optical transitions in monolayer MoSe$_2$ and monolayer WSe$_2$, finding that indirect excitons in the latter correspond to several low-energy optical transitions that so far were attributed to excitons bound to impurities. A possible mechanism that can explain the vanishing polarization in MoSe$_2$ is discussed. Finally, we consider the effect of an out-of-plane electric field, showing that it can reduce the initial polarization of bright excitons due to a Rashba-type coupling with dark excitons.
\end{abstract}
\maketitle

The valley and layer degrees of freedom in transition-metal dichalcogenide semiconductors have appealing pseudospin character, which can be used in applications and explorations of new physical phenomena.\cite{Mak_PRL10,Xiao_PRL12,Cahangirov_PRL12,Wang_NatNano12,Lu_PRL13,Gong_NatComm13,Tongay_NanoLett13,Xu_NatPhys14,Goodfellow_optica14,Mak_Science14} The polarization of excitonic optical transitions provides invaluable information on these degrees of freedom.\cite{Zeng_NatNano12,Mak_NatNano12,Cao_NatComm12,Sallen_PRB12,Kioseoglou_APL12,Jones_NatNano13,Mak_NatMater13,Lagarde_PRL14,Zhu_PNAS14,Jones_NatPhys14,Wang_PRB14,MacNeill_PRL15,Wang_APL15,Wang_2DMater15,Zhang_arXiv15} Compared with typical semiconductors,  electron-hole pairs are strongly bound in these materials due to relatively large effective masses of electrons and holes as well as moderate dielectric constants. Combined with the impeded Coulomb screening in two-dimensional systems, electron-hole pairs can remain bound at room temperature. Inspection of the excitonic luminescence in various  transition-metal dichalcogenides reveals several particular features. Unlike other members of the family, the low-temperature luminescence from monolayer WSe$_2$ has several excitonic peaks closely below the energy of the free exciton ($X_0$).\cite{Jones_NatNano13,Wang_PRB14} Following excitation by a linearly polarized light, only the free exciton retains such polarization. On the other hand, all excitons exhibit some polarization level when excited by a circularly polarized light. When performing the experiment with bilayers, the excitonic transitions retain the polarization of the excitation light.\cite{Jones_NatPhys14,Zhu_PNAS14} Monolayer MoSe$_2$, on the other hand, shows a sharp spectrum merely consisting of free and charged exciton peaks. Both peaks show nearly no polarization regardless of the polarization of the excitation light (linear or circular),\cite{MacNeill_PRL15,Wang_APL15,Wang_2DMater15}   and this behavior persists to bilayer and trilayer MoSe$_2$.\cite{private_comm1,private_comm2}

The purpose of our work is to try elucidating the above phenomena by providing a consistent physical picture. We explain the identity and polarization behavior of optical transitions in monolayer and bilayer transition-metal dichalcogenides (ML-TMDs and BL-TMDs), and shed new light on the interplay between intervalley scattering and electron-hole exchange, and the role of an out-of-plane electric field. The findings are valuable for the interpretation of experimental results, and for elucidating the intrinsic limits for carrying pseudospin information in valleytronics.\cite{Xiao_PRL12,Mak_Science14}

\begin{figure}
    \includegraphics[width=8cm]{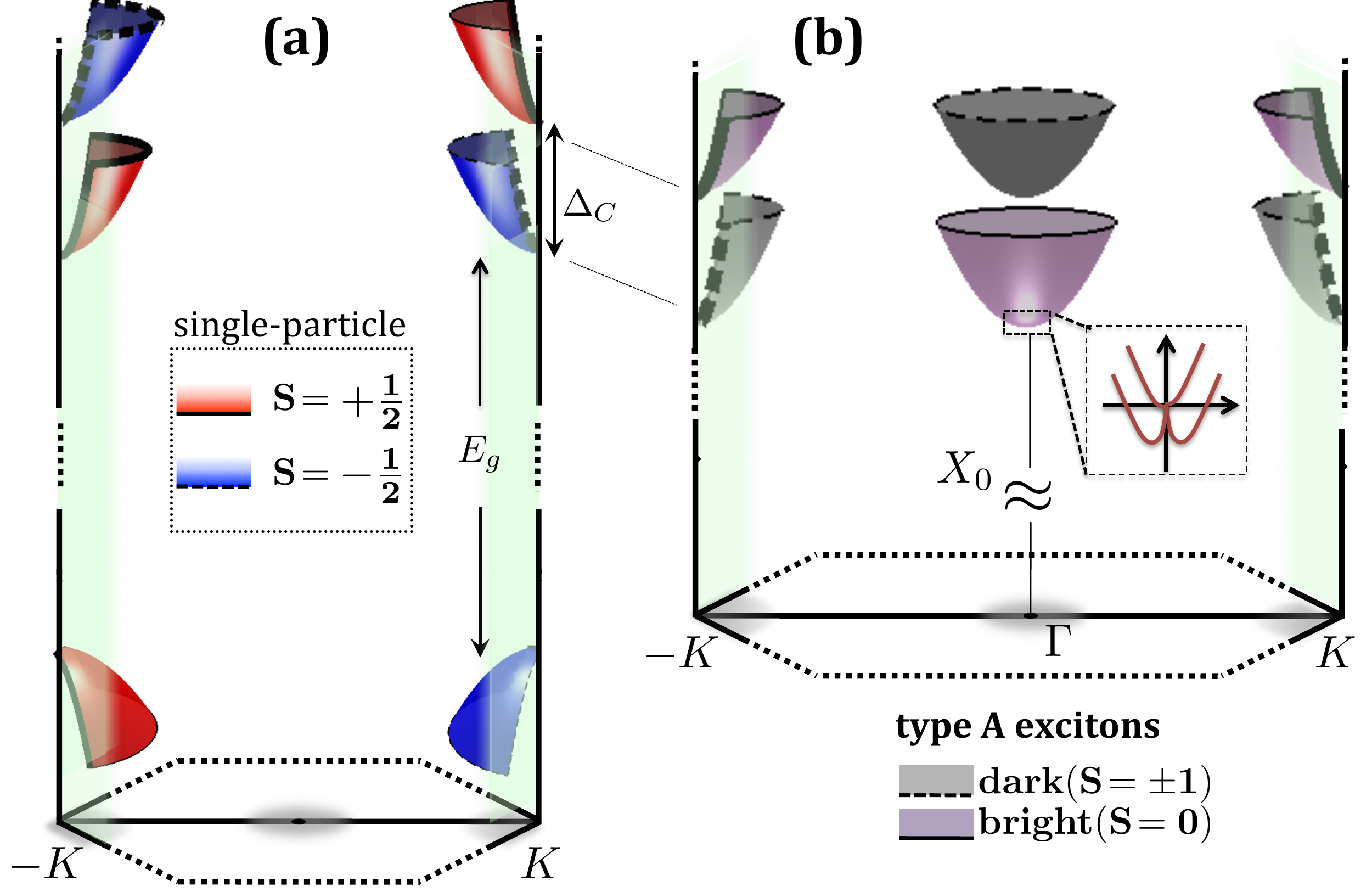}
    \caption{(a) Low-energy band diagrams of electrons and holes in ML-TMDs. The spin-order of conduction bands is that of molybdenum-based TMDs, while the opposite order applies for tungsten-based TMDs.\cite{Liu_PRB13,Kormanyos_2DMater15}  The spin-splitting in the conduction band, $\Delta_c$, is much smaller than the energy gap (see Table~\ref{tab:params}). (b) The resulting bands of type A excitons. The dark-bright exciton splitting is governed by $\Delta_c$. Zone-center bands (direct excitons) are doubly degenerate and denote pairs from $K$ or $-K$ valleys. The electron-hole exchange lifts the degeneracy in the bottom of the direct bright band as shown in the highlighted box.\cite{Yu_NatComm14} Indirect exciton bands in the edge valleys are singly degenerate where each denotes a specific combination of valley and spin for the electron and hole.
        } \label{fig:TMD}
\end{figure}

Figure~\ref{fig:TMD}(a) shows low-energy valleys of electrons and holes in ML-TMDs. Due to the relatively large spin-splitting in the valence band,\cite{Cheiwchanchamnangij_PRB13,Liu_PRB13,Kormanyos_2DMater15,Molina_PRB11,Terrones_SciRef14,Berkdemir_SciRep13,Ataca_JPCC12} we only consider the topmost spin-split band of holes in each valley. For electrons, on the other hand, the spin splitting is much smaller and we consider both spin bands in each valley. Using these electron and hole bands, we classify the so-called type A excitons in ML-TMDs according to their spin and valley. Given that the orbital transition is dipole allowed, excitons are optically active (bright) or inactive (dark)  when the spins of the electron in the conduction band and the missing electron in the valence band are parallel or antiparallel, respectively. In addition, excitons are said to be direct  if the  electron and hole reside in valleys that are centered around the same point in the Brillouin zone. Thus, direct excitons reside in the zone center ($\Gamma$ valleys), while their electron and hole components can be from any part of the Brillouin zone (provided that $k_h - k_e \approx 0$).  In ML-TMDs, a $\Gamma$-valley exciton comprises electron and hole from the $K$ valley (or $-K$ valley).  Similarly, excitons are said to be indirect if the  electron and hole are from opposite  valleys (e.g., electron from $K$ and hole from$-K$). Accordingly,  radiative recombination of an indirect bright exciton involves the  assistance of a phonon or point defect in order to obey or alleviate crystal-momentum conservation, respectively.\cite{Lax_PR61,Cardona_book,Li_PRL10,Li_PRB13}

Using the above classifications,  Fig.~\ref{fig:TMD}(b) shows the resulting low-energy bands of type A excitons in ML-TMDs. The order of bright and dark exciton bands follows the order of the spin-split conduction bands used in Fig.~\ref{fig:TMD}(a).  The direct bright exciton has lower energy than the direct dark exciton and vice-versa for the indirect excitons. This ordering applies for ML-MoSe$_2$, while its reverse applies for ML-WSe$_2$. As will be  explained, this difference can have profound effect on the luminescence. Below, we briefly discuss the excitation of type A excitons and their relaxation to the bottom of the bands. We then focus on the identity and polarization of the ensuing optical transitions.

Light absorption generates primarily direct bright excitons due to their largest optical transition amplitude. The energy degeneracy of the $K$ and $-K$ valleys results in doubly degenerate bands of direct excitons, which for the bright branch are distinguished by the valley index or equivalently by light helicity ($m_l = \pm1$). Accordingly, linearly polarized light excites a valley superposition of the $m_l = \pm1$ states, while circularly polarized light excites one of them. In either case, photoluminescence (PL) experiments show that the $X_0$ peak, which is related to energy-relaxed direct bright excitons, largely retains the polarization of the excitation light in all but SL-MoSe$_2$.\cite{Jones_NatNano13,Lagarde_PRL14,Zhu_PNAS14,Jones_NatPhys14,Wang_PRB14,MacNeill_PRL15,Wang_APL15,Wang_2DMater15,Zhang_arXiv15,private_comm1,private_comm2} The conservation of linear polarization indicates faster intra-valley relaxation of hot excitons compared with the inter-valley scattering of their electron components. The conservation of circular polarization indicates that the intravalley relaxation is also faster than the electron-hole exchange effect. The latter mixes the $m_l = \pm 1$ states.\cite{Maialle_PRB93} While the energy relaxation of hot excitons is not the focus of this study, we mention a few important relaxation channels in Appendix~\ref{app:A}  with emphasis on the interaction of excitons with long-wavelength optical phonons.

\vspace{-3mm}
\subsubsection*{The effect of electron-hole exchange}
\vspace{-3mm}

Following polarized excitation, the luminescence from TMDs can become unpolarized if the timescale for exchange-induced valley mixing is faster than the recombination lifetime. We follow the theory of Yu \textit{et al.},\cite{Yu_NatComm14} and explain the effect of electron-hole exchange on direct bright excitons in TMDs. Away from the $\Gamma$-point, the exchange lifts the energy degeneracy according to $E_0 + E_k \pm J_0k$ where the three terms correspond to edge, kinetic, and exchange energies of direct bright excitons [see highlighted box in Fig.~\ref{fig:TMD}(b)]. The exchange-induced splitting depends linearly on crystal momentum ($k$), where $J_0 \sim 1$~eV$\cdot\AA$.\cite{Yu_NatComm14} In the light cone, where $k \sim 2\pi/\lambda_{light}$, the splitting is of the order of $\sim$1~meV. The resulting oscillations between components of a prepared state can therefore have a period of few ps (Dyakonov-Perel type precession of the valley degree of freedom).\cite{Maialle_PRB93,Dyakonov_JETP71,Tuan_NatPhys14} This timescale is significantly faster than in typical semiconductors due to the tight overlap between the wavefunctions of the electron and hole (e.g., an unscreened exciton extends over areas of  $\sim$1~nm$^2$ in MoS$_2$ vs 100~nm$^2$ in GaAs). The net effect is that polarization of the $X_0$ peak drops if the valley-mixing, induced by the electron-hole exchange, is faster than recombination. Applying a large magnetic field can increase the circular polarization due to the suppression of the exchange-induced valley mixing by lifting the $\Gamma$-point energy degeneracy of direct bright excitons.\cite{MacNeill_PRL15,Wang_2DMater15}

\begin{table}% [h]
%\begin{center}
\caption{Conduction-band spin splitting and pertinent phonon energies in ML-TMDs. Figure~\ref{fig:WSe2}(b) shows the atomic displacements that correspond to these phonon modes. In Koster notation, $A'_1 \rightarrow \Gamma_1$, $E_1^{''}  \rightarrow \Gamma_5$, and $E'_2  \rightarrow \Gamma_6$. \label{tab:params}}
\renewcommand{\arraystretch}{2.0}
\tabcolsep=0.15 cm
\begin{tabular}{c|cccc|c}%{p{0.62in}|p{1.22in}|p{1.22in}}%
 \hline\hline     &   WSe$_2$       &   MoSe$_2$       &   WS$_2$     &   MoS$_2$        &       Ref.                                                          \\
 \hline
$\Delta_c$ (meV)        & -37       & 21                  & -32          & 3                & [\onlinecite{Cheiwchanchamnangij_PRB13}]-[\onlinecite{Kormanyos_2DMater15}]           \\
$E_{A'_1}$ (meV)       & 31         & 30                  & 52           & 51               & [\onlinecite{Molina_PRB11}]-[\onlinecite{Berkdemir_SciRep13}]   \\
$E_{E'_2}$ (meV)       & 31         & 36                  & 45           & 49               & [\onlinecite{Molina_PRB11}]-[\onlinecite{Berkdemir_SciRep13}]   \\
$E_{E_1^{''}}$ (meV)  & 22         & 21                  & 37           & 36               & [\onlinecite{Molina_PRB11}]-[\onlinecite{Berkdemir_SciRep13}]   \\
$E_{K_3}$ (meV)        & 29         & 35                  & 43           & 42               & [\onlinecite{Molina_PRB11}]-[\onlinecite{Ataca_JPCC12}]         \\
\hline\hline
\end{tabular}
%\end{center}
\end{table}

\vspace{-3mm}
\subsubsection*{Photoluminescence in ML-WSe$_2$}
\vspace{-3mm}

\begin{figure}
    \includegraphics[width=8.5cm]{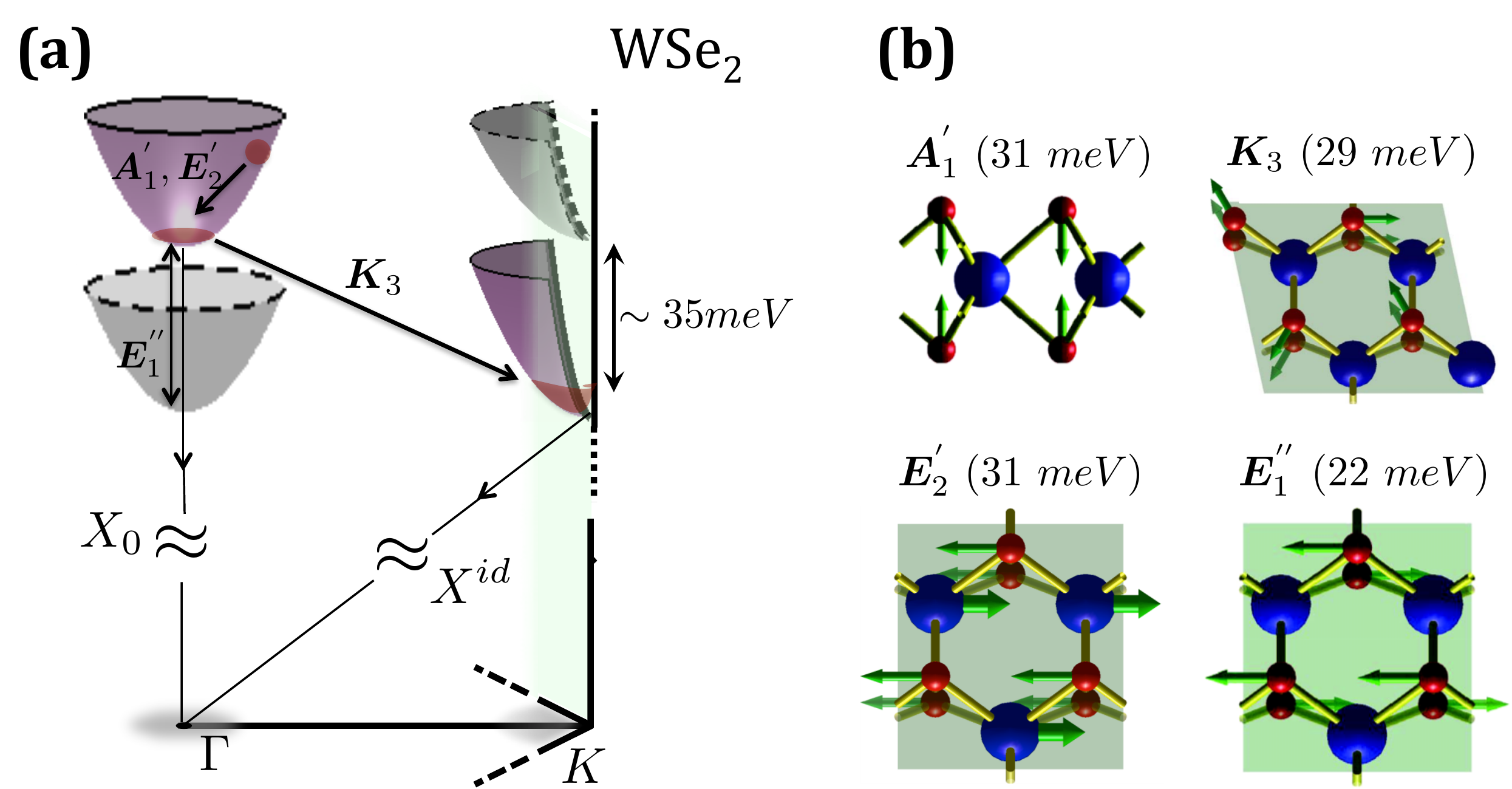}
    \caption{Luminescence process in monolayer WSe$_2$. After energy relaxation to the bottom of the direct bright branch, excitons either recombine or scattered to other branches. Optical phonon modes that participate in the intravalley and intervalley relaxation are indicated, and the corresponding atomic displacements and phonon energies are shown in (b). Direct optical transitions ($X_0$) can retain the excitation polarization, while indirect ones ($X^{id}$) can do so only for circularly-polarized excitation (see text).
        } \label{fig:WSe2}
\end{figure}

In this material, the indirect bright branch has lower energy than the direct bright branch. Figure~\ref{fig:WSe2} shows the energy diagram along with processes that govern the luminescence. Following Table~\ref{tab:params}, ML-WSe$_2$ is the only member in which the dark-bright energy splitting is larger than the energy of the $K$-point phonon needed for intervalley scattering (i.e., in which $|\Delta_c| > E_{K_3}$). Accordingly, energy-relaxed direct bright excitons do not experience phonon bottleneck in ML-WSe$_2$ and eventually become indirect. This physical picture means that optical transitions across the direct gap ($X_0$) represent radiative recombination events just before direct bright excitons turn indirect.  Accordingly, $X_0$ can retain linear polarization  since its luminescence takes place before the intervalley scattering (i.e., the valley superposition is preserved). Similarly, $X_0$ can retain circular polarization if the timescale for intervalley scattering is faster than that for electron-hole exchange. This behavior is indeed corroborated in experiments.\cite{Jones_NatNano13,Wang_PRB14,Yan_arXiv15}

\vspace{-5mm}
\subsubsection*{Photoluminescence  of indirect excitons in ML-WSe$_2$}
\vspace{-3mm}

Figure~\ref{fig:WSe2}(a) shows that in addition to $X_0$, the PL in ML-WSe$_2$ includes optical transitions from indirect bright excitons ($X^{id}$). Whereas their optical transition amplitude is weaker than that of direct bright excitons, they can have evidently larger population at low temperatures. Accordingly, one can expect comparable luminescence from direct and indirect bright excitons at low temperatures. At elevated temperatures, on the other hand, the population of direct and indirect bright excitons becomes comparable (absorption and emission of intervalley phonons have equivalent amplitudes), resulting in a dominant contribution from the $X_0$ transition. Recalling that optical transitions of indirect bright excitons require assistance of phonons or point defects, we can expect several of these peaks. The energy of the highest indirect peak is $X_i^{id} = X_0 - \Delta_c$, and it is associated with an impurity-assisted optical transition mediated by elastic scattering off point defects. The energy of the second highest peak is $X_K^{id} = X_0 - \Delta_c - E_{K_3}$, and it is associated with a phonon-assisted optical transition via zone edge modes with $K_3$ symmetry.\cite{Song_PRL13} The conservation of crystal momentum due to translation symmetry is enabled by virtual intervalley scattering to the light-cone region in the bottom of the direct bright exciton band. Finally, one can expect weaker peaks at larger phonon replicas such as $X_{K+\Gamma}^{id} = X_0 - \Delta_c - E_{K_3} - E_{A_{1}^{'}/E_{2}^{'}}$ which involves emission of a zone-edge and a zone-center phonon modes [see Fig.~\ref{fig:WSe2}(b) for these phonon types]. Here the zone-edge phonon virtually scatters the exciton to the zone center, where it can interact with another long wavelength optical phonon (Raman active modes). Inspection of the parameters of ML-WSe$_2$ in Table~\ref{tab:params} reveals that $X_i^{id}$, $X_K^{id}$ and $X_{K+\Gamma}^{id}$ should appear between $\sim$35~meV and $\sim$100~meV below $X_0$.

%Since the vast majority of literature uses cm$^{-1}$ for phonon energy, I keep it here for the convenience of comparison among them. The unit conversion is 1 cm$^{-1}$=0.124 meV. For more data, see Ref.~\onlinecite{Terrones_SciRef14} for many collections $\Gamma$ phonons of all 4 materials, and  Ref.~\onlinecite{Ataca_JPCC12} for all calculated values.

\vspace{-5mm}
\subsubsection*{Polarization of indirect excitons in ML-WSe$_2$}
\vspace{-3mm}

Indirect excitons do not preserve linear polarization but can keep circular polarization. By definition, the transition of excitons from the direct branch in which they were generated to the indirect branch involves intervalley scattering. As a result, the prepared valley superposition in case of excitation with a linearly polarized light is destroyed. On the other hand, retaining circular polarization is possible due to combination of two factors. Firstly, the holes keep their spin throughout all relaxation processes (until the radiative recombination). Secondly, the indirect exciton bands are not prone to electron-hole exchange since they are singly degenerate.\cite{Song_PRL13}  Previous experiments attributed the optical transitions below $X_0$ in ML-WSe$_2$ to excitons bound to impurities.\cite{Jones_NatNano13,Wang_PRB14} The energy difference from $X_0$ was then associated with the binding energy of the exciton to the impurity. We argue that some of these peaks should be attributed to the indirect bright excitons $X_i^{id}$, $X_K^{id}$ and $X_{K+\Gamma}^{id}$. Not only that their polarizations and energies fit the experiment, our physical picture can explain why these peaks are observed in ML-WSe$_2$ but not in ML-MoSe$_2$. If binding to impurities is the origin, then one is confronted with the fact that similar peaks are not seen in ML-MoSe$_2$.

\vspace{-5mm}
\subsubsection*{Photoluminescence in BL-TMDs}
\vspace{-3mm}
The physical picture presented so far implies that optical transitions associated with $X_0$ are expected to preserve the excitation polarization in BL-TMDs (linear or circular). Contrary to ML-TMDs, bilayers are \textit{indirect} band gap semiconductors in which the interlayer coupling is mostly governed by $p$-orbitals of the chalcogen atoms. Therefore, the coupling only slightly shifts the energy gap of the $K$-point whose electron and hole states are mainly governed by the $d$-orbitals of transition-metal atoms. The result is that ground-state excitons are indirect, and their energy is below that of either direct or \textit{indirect} excitons that comprise electrons and holes of the $K$ and $-K$ valleys. The radiative recombination of direct excitons that govern $X_0$ in bilayers is therefore limited to the time window between their excitation to the direct bright branch and intervalley scattering to the \textit{indirect} branch. A short time window can readily suppress the exchange-induced valley mixing of direct bright excitons.  This physical picture explains the results of several recent experiments with bilayers, in which $X_0$ retained the excitation polarization.\cite{Jones_NatPhys14,Zhu_PNAS14}

%Valley mixing due to electron-hole exchange can reduce the polarization of excitons if the intervalley relaxation is hindered. This would be the case when both electron and hole change their wavefunctions upon exciton scattering from direct to the indirect branches, thereby necessitates a two-phonon process for intervalley scattering. This scenario can even take place in monolayers if the ground-state exciton is made of electron-hole pair from the satellite valleys. Namely, hole states around the $\Gamma$-point and electron states around the local minimum between the $\Gamma$ and $K$ points ($T$ valley). While these satellite valleys have slight higher energies than the $K$ valleys of the valence and conduction bands, the resulting excitons can have very large binding energies due to the heavy masses in these regions: $m_{h_{\Gamma}} > m_0$ and $m_{e_{T}} \sim m_0$.\cite{Kormanyos_2DMater15}

\vspace{-5mm}
\subsubsection*{The anomalous behavior of MoSe$_2$}
\vspace{-3mm}

Contrary to the cases of WS$_2$, WSe$_2$, and MoS$_2$, the PL spectrum of type A excitons in MoSe$_2$ is unpolarized even when the polarized excitation is close to the $X_0$ transition. This anomalous behavior persists in monolayers,\cite{MacNeill_PRL15,Wang_APL15,Wang_2DMater15,Zhang_arXiv15}  bilayers,\cite{private_comm1,private_comm2} and multilayers.\cite{private_comm2} As shown in Fig.~1, the direct bright branch is lower in energy than the direct dark branch in this compound. Therefore, indirect bright excitons should not affect  the PL spectrum due to their small population and weak optical-transition amplitude. It is tempting to justify the vanishing polarization of direct bright excitons by arguing that the valley mixing, induced by electron-hole exchange, is faster than recombination or intervalley scattering.  However, time-resolved PL experiments reveal recombination times of the order of a few ps,\cite{Wang_APL15} similar to those found in other TMDs.\cite{Lagarde_PRL14,Wang_PRB14} In addition, a recent PL experiment in ML-MoSe$_2$ showed that when the optical excitation is from the bottom spin-split valence band to the conduction band (type B excitons), the luminescence of these higher-energy excitons remains polarized.\cite{Zhang_arXiv15} This observation suggests that the electron-hole exchange of type B excitons is not efficient within the time window between their excitation and energy relaxation to the type A branch. Importantly, the magnitude of electron-hole exchange of type A excitons should be similar to that of type B excitons on account of their similar symmetries and atomic orbitals. All these facts imply that enhanced effect of the electron-hole exchange in MoSe$_2$ compared with other TMDs is not likely to be the cause for the PL's vanishing polarization.

We conjecture that the anomalous behavior in MoSe$_2$ can be reasoned by polaron-induced coherent coupling between the direct branches of bright and dark excitons.  By inspecting the transformation properties of these excitons, they can only be coupled by the $E^{''}_1$ phonon mode whose corresponding atomic displacement is shown in Fig.~2(b).\cite{Song_PRL13} Of all TMDs, the energy of this phonon resonates with the conduction-band spin-splitting only in MoSe$_2$ ($\Delta_c \simeq E_{E_1^{''}}$; See Table 1). A possible outcome is a polaron-induced coupling that drives coherent  oscillations between the direct-bright and direct-dark excitons, where on each return to the bright branch the helicity changes sign (Rabi oscillations in a two-level system where each level is doubly degenerate). This process can be accountable for the vanishing polarization if the electron-phonon coupling amplitude is strong enough to support sub-one ps oscillations. The physics should be similar in ML-MoSe$_2$ and multilayer-MoSe$_2$ since the $\Gamma$-point phonon energies and $K$-point splitting of the conduction band hardly change with the number of layers.

%A second explanation that can give rise to the vanishing polarization in MoSe$_2$ is a phonon bottleneck that impede intervalley exciton transitions.

\vspace{-5mm}
\subsubsection*{`Brightening'  dark excitons in ML-TMDs}
\vspace{-3mm}
Spin flips of the electron or hole component  induce  transitions between bright and dark excitons.\cite{Song_PRL13} The other way to couple dark and bright excitons in monolayers, where space inversion is not respected, is via the presence of an out-of-plane electric field (e.g., by a gate voltage or charged impurities in the substrate).

To correctly quantify the resulting Rashba-type coupling between bright and dark components of direct excitons, we follow their transformation properties according to the group theory analysis in Ref.~[\onlinecite{Song_PRL13}]. The resulting Hamiltonian matrix of direct excitons reads,
\begin{equation}
H =  \left( \begin{array}{cc} H_b & H_R \\ H_R^{\ast} & H_d  \end{array} \right). \label{eq:H}
\end{equation}
The upper diagonal block belongs to bright excitons,\cite{MacNeill_PRL15,Yu_NatComm14}
\begin{equation}
H_b =  \frac{\hbar^2k^2}{2M}\mathcal{I} + J_0k(\cos{2\theta} \sigma_x + \sin{2\theta}\sigma_y) + \eta_b B_z \sigma_z, \label{eq:Hb}
\end{equation}
where we have incorporated  the possibility of an out-of-plane magnetic field in addition to exchange and kinetic terms. $M$ is the free exciton mass (about twice the electron mass in TMDs), $\mathbf{k} = k(\cos{\theta}, \sin{\theta})$, and $\eta_b \approx 2\mu_B$ due to valence-band orbitals of the transition-metal atoms. Working in this subspace (in which basis states transform as $\Gamma_6$), one can quantify the competition between electron-hole exchange and the magnetic field. The lower diagonal block in Eq.~(\ref{eq:H}) is of dark excitons,
\begin{equation}
H_d =   \Delta_{bd} + \frac{\hbar^2k^2}{2M}\mathcal{I}  + (J_d + \eta_d B_z)\sigma_z, \label{eq:Hd}
\end{equation}
where $\Delta_{bd} \approx \Delta_c$ is the bright-dark energy splitting (we consider the energies of the $k=0$ states in the bright branch as the reference level). The last term includes short-range exchange interaction as well as Zeeman terms. The existence of exchange-induced splitting is supported by the fact that the basis states in the dark branch transform as one-dimensional rather than two-dimensional irreducible representations  ($\Gamma_3$ and $\Gamma_4$). The Zeeman term of dark excitons,  $\eta_d \approx 4\mu_B$, has contributions from the spin and valence-band orbitals of the transition-metal atoms. Finally, the off-diagonal block in Eq.~(\ref{eq:H}) represents the Rashba coupling between bright and dark excitons,
\begin{equation}
H_R =  \alpha_R k E_z \left( \begin{array}{cc} \exp(-i\theta)  & \exp(-i\theta)  \\  - \exp(i\theta)  & \exp(i\theta)   \end{array} \right), \label{eq:HR}
\end{equation}
where $E_z$ is the out-of-plane electric field and $\alpha_R$ is the Rashba coefficient (see Appendix~\ref{app:B} for its analytical expression).

To see the coupling between bright and dark excitons, it is sufficient to diagonalize Eq.~(\ref{eq:H}) by neglecting the magnetic and exchange terms (their energy scales are smaller than the bright-dark energy splitting).  The new dark states are `brighten' by the E-field according to,
\begin{equation}
\widetilde{\Psi}_{d1,d2} \approx   \Psi_{d1,d2}  + \frac{\alpha_R k E_z}{\Delta_{bd}} \left(  \Psi_R e^{-i\theta} \pm  \Psi_L e^{i\theta} \right).   \label{eq:Psi}
\end{equation}
where $\Psi_{R,L}$ are the basis states of bright excitons (right or left helicity), and $\Psi_{d1,d2}$ are the basis states of dark excitons without the E-field. %Their respective energies become
%\begin{equation}
%E_d =   \Delta_{bd} + \frac{\hbar^2k^2}{2M} - \frac{ (\alpha_R E_z k)^2}{\Delta_{bd}}.   \label{eq:E}
%\end{equation}
Similar expressions can be derived for bright excitons. The E-field induced `brightening' of dark excitons (and `darkening' of bright excitons) becomes evident away from the zone center. This mixing degrades the attainable valley polarization upon excitation above resonance in ML-TMDs.

\vspace{-3mm}
\subsubsection*{Concluding remarks}
\vspace{-3mm}

We have attempted to elucidate many of the features of optical transitions in transition-metal dichalcogenides. An important effect is that energy-relaxed direct bright excitons, which correspond to the so-called $X_0$ peak, can retain the linear or circular polarization of the excitation light when two intrinsic conditions apply. The first one is that their energy is above that of indirect bright excitons, and the second one is that intervalley transitions are not impeded by a phonon bottleneck. These conditions apply in monolayer WSe$_2$ for which we have associated several of the observed optical transitions with indirect bright excitons, and explained their polarization behavior. These conditions can also be applied in multilayers if the recombination lifetime of direct bright excitons is limited by an ultrafast time window between excitation and intervalley scattering to the low-energy indirect gap.

The anomalous vanishing polarization in monolayer and multilayer  MoSe$_2$ remains an open question. We have alluded to a subtlety in the band diagram of excitons in MoSe$_2$ by noticing that the energy splitting between dark and bright exciton branches resonates with the energy of the phonon that couples these branches. This resonance condition, not met in other transition-metal dichalcogenides,  can result in vanished polarization due to polaron-induced Rabi oscillations between dark and bright excitons.

Finally, we have presented the Rashba coupling between dark and bright  components of direct excitons in the presence of an out-of-plane electric field. The coupling is amplified away from the zone center, and therefore can affect the initial polarization degree of excited electron-hole pairs.

%An open question, not dealt with in this work, concerns the optical properties of charged excitons. Given the strong phonon signatures on optical transitions in chalcogen-based materials,\cite{Fomin_PRB98} the luminescence of charged excitons in transition-metal dichalcogenides could be a signature of plasmon-phonon coupling rather than of pure trion states (e.g., one hole with two electrons).

\acknowledgments{We are grateful for many useful discussions with Xiaodong Xu who proposed this topic. We are thankful to Aubrey Hanbicki, Aaron Mitchell Jones, Jie Shan, Kin Fai Mak, and Nick Vamivakas for sharing valuable experimental results prior to their publication. Finally, we are grateful to Wang Yao for illuminating discussions on the effect of electron-hole exchange. This work was supported by the National Science Foundation under Contracts No. DMR-1503601, No. ECCS-1231570, No. DMR-1124601, and the Defense Threat Reduction Agency under Contract No. HDTRA1-13-1-0013.}

\appendix

\section{Energy relaxation of hot excitons} \label{app:A}
The energy relaxation is mediated by interaction of excitons with other excitons, free-charges, impurities, or phonons. Relaxation via exciton-exciton or exciton-plasmon scattering becomes relevant for  high-intensity excitation,\cite{Kumar_PRB14,You_NatPhys15} or when the background-charge density is large.\cite{Pezzoli_PRB13} Relaxation can also be governed by the displacement of point-defect impurities,\cite{Henry_PRB77}  similar to the case of interface states in quantum dots.\cite{Sercel_PRB95,Schroeter_PRB96} Here we will focus, however, on the intrinsic relaxation in high-quality crystals subjected to weak excitation and low background density of free charges. In this case, emission of long-wavelength optical phonons is the most relevant relaxation path, and it can take place in less than one ps if only a handful of phonon emissions are needed in order to reach the bottom of the type A exciton branch.

From symmetry,  only the Raman active modes $A_{1}^{'}$ and $E_{2}^{'}$, shown in Fig.~\ref{fig:WSe2}(b), are involved in intravalley energy relaxation ($\Gamma_1$ and $\Gamma_6$ in Koster notation).  The $A_{1}^{'}$ mode corresponds to thickness fluctuations while the $E_{2}^{'}$ mode corresponds to in-phase motion of the chalcogen atoms in the plane of the layer. The phonon-exciton interaction in both cases have long-range and short-range components. The long-range interactions are due to Frohlich-type coupling between the excitons and the in-plane macroscopic electric field generated by both atomic displacements in the long-wavelength limit. The exciton does not change its polarization state since both long-range interactions transform as the identity irreducible representation. The interaction of neutral excitons with the long-range macroscopic field is expected to be weak due to similar but opposite-sign contributions from the electron and hole components.

The short-range interactions of excitons with the $A_{1}^{'}$ and $E_{2}^{'}$ phonon modes have opposite polarization effects. The exciton does not change its polarization state by the short-range interaction with  $A_{1}^{'}$ modes, since the corresponding atomic displacement transforms as the identity irreducible representation. The amplitude of this short-range interaction is governed by the deformation potential of the thickness fluctuations.\cite{Fivaz_PR67,Kaasbjerg_PRB14} On the other hand, the exciton changes its polarization state due to the short-range interaction with $E_{2}^{'}$ modes since their atomic displacement complies with the transformation properties of off-diagonal Pauli matrices ($\sigma_x$ and $\sigma_y$).\cite{Song_PRL13} This scattering therefore flips the helicity (valley state) of bright excitons in case of a circularly-polarized prepared state, and destroys the polarization in case of a linearly-polarized prepared state. This behavior was indeed corroborated in a recent experiment.\cite{Chen_NanoLett15}

% However, the effect is expected to be weaker than that of the long-range one which conserves the polarization state. Specifically, changing the polarization of an exciton by long-wavelength phonons requires high-order  contributions. For example, it can be mediated by two phonons with overall $\Gamma_6$ symmetry such that one interacts with the electron and the other with the hole. Alternatively, it can be a confluence of one phonon and Coulomb interactions of the form $V(\mathbf{r}_1-\mathbf{r}_2)* H_{{E_{2}^{'}}}$, where the exciton's nonlocal Coulomb part mediates the valley flip.

\section{Rashba coefficient } \label{app:B}
Using $\mathbf{k}\cdot \mathbf{p}$ state expansion, the Rashba coefficient can be written as
\begin{eqnarray}
\alpha_R &\approx& \frac{ie \hbar}{m}  \left[ \frac{ \langle \Psi_{6,x}| z |\Psi_{5} \rangle  \langle \Psi_{5}| p_y |\Psi_{3} \rangle  }{\varepsilon_6-\varepsilon_{5}}  \right.
 \nonumber \\ & \, & \qquad \qquad  \left. + \frac{ \langle \Psi_{6,x}| p_y |\Psi_{2} \rangle  \langle \Psi_{2}| z |\Psi_{3} \rangle}{\varepsilon_6-\varepsilon_{2}}    \right]  \label{eq:alpha}
\end{eqnarray}
where $\Psi_i$ represent the basis states of irreducible representations $\Gamma_i$ in Koster notation. $\Gamma_6$ corresponds to bright excitons where $2\Psi_{6,x} = \Psi_{6,L}+\Psi_{6,R}$. The dark excitons are represented by $\Gamma_{3}$.  Equivalent expressions can be derived with the help of $\Psi_4$ and $\Psi_{6,y}$ (the other pair of dark and bright states).  The symmetry allowed intermediate states are represented by $\Gamma_{2,5}$ irreducible representations. Type B dark excitons, which belong to $\Gamma_5$, seem to contribute the most since the energy denominator in Eq.~(\ref{eq:alpha}) becomes the valence-band spin splitting, which is evidently smaller than for all other remote bands with $\Gamma_{2,5}$ symmetries.  The expression for $\alpha_R$ suggests that it scales linearly with the spin mixing of the conduction band components of exciton states. The reason is that  initial ($\Psi_{3,4}$) and final ($\Psi_6$) states have opposite spins for conduction components.  In deriving the expression for $\alpha_R$,  we have used $\langle \Psi_{6,x} \pm i \Psi_{6,y}|...| \Psi_4\rangle\sim \Gamma_5 \sim (x\mp iy)z$ and $\langle \Psi_{6,x} \pm i \Psi_{6,y}|...| \Psi_3\rangle \sim \Gamma_5 \sim (\pm x- iy)z$. Finally, we note that  $\alpha_R$ can be extracted from ab-initio techniques by fitting a $(\alpha_R E_z k)^2/\Delta_{c}$ component in the conduction band energy dispersion as a function of $E_z$.

\end{document}